\documentstyle[12pt,epsfig]{article}
\def\lapp{{\ \lower 0.6ex \hbox{$\buildrel<\over\sim$}\ }}
\def\gapp{{\ \lower 0.6ex \hbox{$\buildrel>\over\sim$}\ }}
\def\half{{\textstyle{1\over 2}}}
\begin{document}
\begin{titlepage}
\vspace*{-1cm}
\begin{flushright}
DTP/95/64   \\
July 1995 \\
\end{flushright}
\vskip 1.cm
\begin{center}

{\Large\bf Higher-order Coulomb Corrections to the Threshold
$e^+e^-\to W^+W^-$ Cross
Section}
\vskip 1.cm
{\large  V.S.~Fadin}
\vskip .2cm
{\it Budker Institute for Nuclear Physics and Novosibirsk State
University \\ 630090 Novosibirsk, Russia}\\
\vskip   .4cm
{\large  V.A.~Khoze, A.D.~Martin}
\vskip .2cm
{\it Department of Physics, University of Durham \\
Durham DH1 3LE, England }\\
\vskip   .4cm
and
\vskip .4cm
{\large  W.J.~Stirling}
\vskip .2cm
{\it Departments of Physics and Mathematical Sciences, University
of Durham \\
Durham DH1 3LE, England }\\
\vskip 1cm

\end{center}

\begin{abstract}
The QED Coulomb correction is one of the most important
corrections
to the $e^+e^- \to W^+W^-$ total cross section near threshold.
We calculate these corrections through second order, and discuss
the implications for extracting $M_W$ from a threshold cross
section
measurement at LEP2. Analytic expressions are derived in various
kinematic limits.
\end{abstract}
\vfill
\end{titlepage}

\newpage

\section{Introduction}
\label{sec:intro}
The magnitude of the $e^+e^-\to W^+W^-$ cross section near
threshold at
LEP2 provides a powerful method for measuring $M_W$ at the LEP2
collider, see for example Ref.~\cite{WJS}. It is sufficient to
make a
single measurement of the cross section at the `collision energy
of maximal
sensitivity', which is approximately 0.5~GeV above the nominal
$\sqrt{s} = 2 M_W$ threshold. The method relies on a
high-precision
calculation of the cross section as a function of $M_W$, and it
is
therefore important that higher-order  corrections are under
control. The QED Coulomb correction is one of the most important
of
these, especially at threshold. Note that since Coulomb physics
---
which is associated with large space-time intervals --- is so
different
from the other radiative effects, it is possible to study its
impact
separately.

As was emphasized in Refs.~\cite{FKM3,FKMC} (see also
\cite{BARDIN}), the
Coulomb corrections to the total $e^+e^-\to W^+W^-$ cross
sections
are relatively small because the instability effects {\it mask}
the Coulomb singularity. In fact the corrections are numerically
more
important in the case of differential distributions, for example
of the
$W$ invariant mass or momentum \cite{WJSVAK}, but even so their
exact
(i.e. all-orders) calculation appears to be unnecessary for the
level of
precision required at LEP2.

The first-order Coulomb correction for the (off-shell)
 $W^+W^-$ total cross section
was first  presented in Ref.~\cite{FKM3,FKMC}, and its
quantitative effects
were discussed in Refs.~\cite{BARDIN,WJS}. It was shown that the
correction is
positive in the threshold region, attaining a maximum value of
approximately $+6\%$ at threshold. In terms of using the size
of  the measured threshold cross section to measure $M_W$, the
inclusion
of the first-order Coulomb correction is equivalent to a shift in
$M_W$
of order $100$~MeV. In view of the target LEP2 precision on $M_W$
of
better than $\pm 50$~MeV, it is clearly important to investigate
the
size of the higher-order contributions.

In this paper we present a quantitative analysis of these
Coulomb contributions,
and in particular their impact on the threshold $W^+W^-$ cross
section.
We do not include initial-state radiation or `hard' radiative
corrections, but their effects would be straightforward to
include,
at least in principle (see for example Ref.~\cite{BEENDENN}).
In the following section we recall the results of
Ref.~\cite{FKMC}
for the higher-order Coulomb contributions to the $W^+W^-$ cross
section, and derive an explicit expression for the $O(\alpha^2)$
contribution. In Section~\ref{sec:closed} we obtain a closed
formula for the
$O(\alpha^2)$ correction which is valid in the non-relativistic
limit,
and in Section~\ref{sec:numer} we derive numerical results for
the
first- and second-order Coulomb corrections over the range of
LEP2
energies. Some brief conclusions are presented in
Section~\ref{sec:conc}.

\section{Higher-order Coulomb effects}
\label{sec:coulomb}
The appropriate formalism for analysing the all-orders Coulomb
effects
in the production of a pair of heavy unstable particles, based on
the
technique of non-relativistic Green's functions, has been known
for some
time, see Refs.~\cite{FK,FADKHO}.
It was developed substantially because of the requirements of $t
\bar t$
threshold production physics \cite{STRASSLER,FKM1,FKK,KUHN}.
The application to unstable $W$ boson production was first
considered
in Ref.~\cite{FKMC}.
A simplified prescription for incorporating high-order Coulomb
corrections in this case has been proposed in Ref.~\cite{BARDIN},
based
on the direct substitution of the one-loop off-shell correction
into the
exact all-orders on-shell result \cite{SOMMERFELD}. Although this
procedure may provide some qualitative understanding of the size
of the
higher-order contributions, it cannot be justified on theoretical
grounds.    In particular, it certainly cannot be applied at or
just
below threshold energies, $E \approx \sqrt{s}-2M_W \leq 0$.
More generally, the first-order
correction is related only to the {\it real} part of the Coulomb
factor
$f(\vec{p},E)$ (see \cite{FADKHO,FKM1,FKMC}), whereas at
higher-orders
both the real and imaginary parts contribute.\footnote{Recall
that for
stable particles the imaginary part of the one-loop correction
contains
an infra-red divergence which is related to the unobservable
Coulomb
phase. This singularity leads to some specific consequences for
the case
of unstable $W$ bosons \cite{FKM1,FKMC}.  In some sense, the
remnant
of the Coulomb phase could be important in differential
distributions,
see for example the discussion in Ref.~\cite{MELYAK2}.}
In what follows, we will quantitatively
compare results for the Coulomb corrections to the total $W^+W^-$
cross section obtained from
the `exact' non-relativistic all-orders prescription of
Ref.~\cite{FKMC}
with the approximate procedure of Ref.~\cite{BARDIN}.

Note that for energies $E \gg \Gamma_W$ the modification of the
Coulomb contribution to the total cross section arising from the
instability of the $W$ bosons is, at most, of relative order
$O(\alpha \Gamma_W/E)$. The modifications due to all other
final-state QED interaction effects are cancelled in the total
cross section, up to terms of relative order $\alpha
\Gamma_W/M_W$ \cite{FKM1,FKM2,FKM3} (see also \cite{MELYAK1}).

The all-orders Coulomb contribution $\delta_C$ to the radiative
correction
to the $e^+e^- \to W^+W^-$ off-shell Born cross section can be
written as
\begin{equation}
\label{deltadef}
1 + \delta_C = \vert f(\vec{p},E) \vert^2 ,
\end{equation}
where the Coulomb enhancement  factor $f(\vec{p},E)$
 is given by \cite{FADKHO,FKK,FKMC}
\begin{equation}
\label{fdef}
f(\vec{p},E) = 1 + \alpha \sqrt{s} \kappa \int_0^1 dx\;
{ x^{-\alpha\sqrt{s}/4\kappa} \over
\kappa^2(1+x)^2 + p^2 (1-x)^2 } ,
\end{equation}
with
\begin{equation}
p^2 ={1 \over 4s}\; \left[
s^2-2s(s_1+s_2)+(s_1-s_2)^2 \right]
\end{equation}
\begin{equation}
\kappa = \sqrt{-M_W(E+i \Gamma_W)} \equiv p_1 -i p_2
\end{equation}
\begin{equation}
p_{1,2} = \left[ \half M_W \left(
\sqrt{E^2 + \Gamma_W^2} \mp E
\right)\right]^{1\over 2}
\end{equation}
\begin{equation}
E = {s - 4M_W^2 \over 4 M_W}
\end{equation}
As discussed in Ref.~\cite{FKMC}, the representation (\ref{fdef})
is convergent for all values of $E$, both above ($E>0$) and below
($E<0$) the nominal $W^+W^-$ threshold. Expanding
$f(\vec{p},E)$ as a power series in $\alpha$ up to $O(\alpha^2)$
terms one finds
\begin{equation}
f(\vec{p},E) \approx 1 + {\alpha\sqrt{s}\over 4 i p}\ln D
 + {\alpha^2 s \over 16 i p \kappa} \; \int_0^1 {dx \over x}\;
\ln \left( { 1 + x D \over 1 + x/D }   \right) ,
\label{fnnlo}
\end{equation}
with
\begin{equation}
D = {\kappa + i p \over \kappa - i p } .
\end{equation}
It is straightforward to verify that the real part  of the second
term on the right-hand side in Eq.~(\ref{fnnlo}) coincides with
the $O(\alpha)$ correction calculated in Refs.~\cite{FKM3,FKMC}:
\begin{eqnarray}
\vert f\vert^2 &\approx& 1 +
2\; \mbox{Re}\, \left( {\alpha\sqrt{s}\over 4 i p}\ln D\right) +
O(\alpha^2)
\nonumber \\
& =& 1 + {\alpha\sqrt{s}\over 4  p}
\left[ \pi -2\; \mbox{arctan}\left({\vert\kappa\vert^2 -p^2
\over
 2 p\; {\rm Re}(\kappa ) }  \right)\right] + O(\alpha^2).
\label{fnlo}
\end{eqnarray}
Eqs.~(\ref{fdef},\ref{fnnlo}) demonstrate explicitly how the
Coulomb
singularity is screened by instability effects. In fact the
expansion
parameter of the Coulomb series is effectively $ \alpha \pi
\sqrt{s}
/ (2 \vert \kappa \vert )$, which attains a maximum value of
$ \alpha \pi \sqrt{M_W / \Gamma_W}\approx {1\over 7} $
at the exact threshold ($E=0$),\footnote{It is interesting to
note that this
effective QED expansion parameter is of the same order as the
strong
coupling constant.}
in contrast to the expansion parameter
$ \alpha \pi \sqrt{s} /(2 p)$ of the stable $W$ case.
In particular, in the limit $p \ll \vert\kappa\vert$, which
corresponds
to the lower $W$ momentum tail, we have explicitly
\begin{eqnarray}
f(\vec{p},E)  &=&  1 -4 \sum_{n=1}^\infty {(-1)^n \over 2^{2n}
(n-1)! }
\left( {\alpha \sqrt{s}\over   \kappa} \right)^n \int_0^1 dx\;
{ \ln^{n-1}x \over(1+x)^2} \nonumber \\
&=& 1 + {\alpha \sqrt{s} \over 2 \kappa} + {\alpha^2 s \over 4
\kappa^2}
\ln 2 + O(\alpha^3) .
\end{eqnarray}

\section{Closed formula in the non-relativistic limit}
\label{sec:closed}
In Refs.~\cite{FK,FADKHO} a general approach has been proposed
which allows
one to obtain a closed formula for the dominant contribution to
the
all-orders Coulomb correction to the total unpolarized cross
section for the
production of a pair of heavy unstable particles.
The application of this to the case of $e^+e^-\to W^+W^-$ has
been  discussed
in Ref.~\cite{FKMC}.
The Coulomb effects are incorporated through the imaginary part
of the
Green's function $G_{E + i \Gamma_W}(\vec{r}= 0,\vec{r'}= 0)$ of
the
Schrodinger equation for the interacting $W^+W^-$ system.
The result for the total $W^+W^-$ cross section is
\begin{equation}
\label{nonrel}
\sigma(s) \; = \;    C\left[
 {2 p_2 \over \sqrt{s}} \; +\;
\alpha \arctan{p_2\over p_1}
\; + \; {\alpha^2\over 2} \zeta(2)\; {p_2 \over \sqrt{E^2 +
\Gamma_W^2} }
\; + \; O(\alpha^3) \right]\; ,
\end{equation}
where $\zeta(2) = \pi^2/6$.
The factor $C$ is related to the expansion of the off-shell Born
cross section
$\sigma_0(s,s_1,s_2)$  \cite{MUTA}
in the non-relativistic region $\beta = 2 p/\sqrt{s}  \ll 1$
(where
$p^2 \simeq (\sqrt{s} - \sqrt{s_1} - \sqrt{s_2}) M_W$),
\begin{equation}
\label{expansion}
\sigma_0 = C \beta + O(\beta^3) \; ,
\end{equation}
which, when folded with the usual Breit-Wigner factors to give
the total
Born cross section,
\begin{equation}
\sigma_B(s) = \int_0^s d s_1 \int_0^{(\sqrt{s}-\sqrt{s_1})^2}
ds_2\;
\rho(s_1)\rho(s_2)\; \sigma_0(s,s_1,s_2) \; .
\label{sigborn}
\end{equation}
leads to the first term on the right-hand side of (\ref{nonrel}),
\begin{equation}
\sigma_B(s) \simeq C\;  {2 p_2 \over \sqrt{s} } \; .
\end{equation}
Note that the Breit-Wigner factors in (\ref{sigborn}) are defined
by
\begin{equation}
\rho(s) = {1\over \pi} {\Gamma_W\over M_W}\; {s\over
(s-M_W^2)^2 + s^2 \Gamma_W^2/M_W^2 }
 \; .
\label{rho}
\end{equation}

The higher-order terms in the expansion (\ref{expansion})
give a  negative correction to the leading $\sim \beta$
behaviour, see for example Ref.~\cite{WJS}. The net effect is a $
O(\Gamma_W/M_W)$
reduction of $\sigma_B(s)$ compared to the zero-width
approximation,
(see, for example, Refs.~\cite{BIGI,FK2})\footnote{Note that in
practice the role
of these $\Gamma_W/M_W$ effects --- which are related to the
large
`intrinsic' momenta of the $W$ bosons --- is likely to be
suppressed
by experimental cuts on the $\sqrt{s_i}$, which are necessary in
order
to separate the $W^+W^-$ events from the non-$W^+W^-$
background.}
\begin{equation}
\label{bornplus}
\sigma_B(s) \simeq C \; {2 p_2 \over \sqrt{s} } \;  -\;
 a {\Gamma_W \over M_W}\;  + \; \ldots \; .
\end{equation}
An explicit numerical calculation (see below) gives $C \approx
54$ pb and $a \approx 80$~pb.

Finally, it is straightforward to show that for $E \gg \Gamma_W$
the
right-hand  side of Eq.~(\ref{nonrel})
reproduces the expansion of the Coulomb factor for stable
particles
\cite{SOMMERFELD},
\begin{equation}
\label{stable}
\vert\psi(0)\vert^2 = {Z \over 1 - \exp(-Z) }  = 1 + {Z \over 2}
+ {Z^2 \over 12} + \ldots \; ,
\end{equation}
with $Z = \alpha \pi /\beta$. The modification caused by the $W$
instability
is important only for $E \lapp \Gamma_W$. Far below threshold, $E
< 0$
with $\vert E \vert \gg \Gamma_W$, Eq.~(\ref{nonrel}) becomes
\begin{equation}
\label{farbelow}
\sigma(s) \simeq
  {C \Gamma_W \over 2 \sqrt{ \vert E \vert M_W} } \;   \left[
 1 +
\alpha \sqrt{M_W \over \vert E \vert}
+ {\alpha^2\over 2} \zeta(2) {M_W \over \vert E \vert} + \ldots
\right]\; .
\end{equation}

\section{Numerical results}
\label{sec:numer}

In this section we present the numerical results of a calculation
of the $O(\alpha)$ and $O(\alpha^2)$ Coulomb corrections to the
total
$e^+e^- \to W^+W^-$ cross section near threshold, i.e.
\begin{equation}
\sigma(s) = \int_0^s d s_1 \int_0^{(\sqrt{s}-\sqrt{s_1})^2}
ds_2\;
\rho(s_1)\rho(s_2)\; \sigma_0(s,s_1,s_2) \; [ 1 +
\delta_C(s,s_1,s_2) ]
\; ,
\label{sigcoul}
\end{equation}
with $1 + \delta_C = \vert f\vert^2$ where $f$ is given by
Eq.~(\ref{fnnlo}).
Our calculation is an extension of that presented in
Ref.~\cite{WJS},
which incorporated the $O(\alpha)$ Coulomb correction, and all
parameters
are identical except that here we use an updated value for the
$W$ mass,
$M_W = 80.41$~GeV/c$^2$ \cite{CDFMW}.

Fig.~1 shows  the $O(\alpha)$ and $O(\alpha^2)$ Coulomb
corrections
normalized to the Born cross section, as a function of the
$e^+e^-$ collider
energy $\sqrt{s}$.  The position of the nominal threshold
$\sqrt{s} =
2 M_W$ is marked. We see that the energy dependence of the first
and second
order corrections is similar --- both attain a  maximum just
below
 threshold of approximately  $+6\%$ and $+0.2\%$ respectively.
In Ref.~\cite{WJS} it was shown that at threshold a change in the
cross section of $\Delta \sigma$ was equivalent to a shift
in the $W$ mass of
\begin{equation}
\Delta M_W \; = \; 17\ {\rm MeV}  \ \cdot \  \left[ {\Delta
\sigma \over
\sigma } \times 100\% \right] \; .
\end{equation}
Evidently, the inclusion of the second-order Coulomb correction
is equivalent to  $\Delta M_W = 3.4$~MeV, which is negligible
in comparison to the anticipated precision on $M_W$ using the
threshold cross section measurement method.

How do these results compare with the analytic approximations
derived
in the previous section? The expansion of Eq.~(\ref{nonrel}) at
$E=0$ is
\begin{equation}
\label{threshold}
\sigma(s = 4 M_W^2) \; = \;
C\;   \sqrt{\Gamma_W \over 2 M_W} \; \left[
 1 \; + \; {\overline{X} \over 2} \; + \;
{\overline{X}^2 \over 6} \; + \; \ldots
 \right]\; ,
\end{equation}
where
\begin{equation}
\label{a22}
\overline{X} \;  = \;
\alpha \pi     \sqrt{M_W \over 2 \Gamma_W} \;  \approx 0.1005 \;
\; .
\end{equation}
This gives first- and second-order corrections of $+5\%$ and
$+0.17\%$ respectively, in good agreement with the exact result.
In fact the difference between these values and the `exact
values' is due simply to the negative $O(\Gamma_W/M_W)$
corrections to the Born cross section (\ref{bornplus}) which are
not taken into account in (\ref{threshold}).    Note also from
Fig.~1 that the first and second order corrections decrease
rapidly below threshold, consistent with the analytic result for
$ E \ll -\Gamma_W$ of Eq.~(\ref{farbelow}).

Returning to the expression for $f$ in Eq.~(\ref{fnnlo}), we see
that the second-order correction contains contributions from both
the real and imaginary parts of the first-order contribution to
$f$, and from the real part of the second-order contribution.
Schematically, from (\ref{deltadef}) and (\ref{fnnlo}) we have
\begin{eqnarray}
\delta_C^{(1)} & \leftarrow & 2 f_1^R   \nonumber \\
\delta_C^{(2)} & \leftarrow & 2 f_2^R  +  (f_1^R)^2 + (f_1^I)^2
\; .
\end{eqnarray}
The following table gives the corresponding breakdown of the
second-order
correction  (in $\%$) at several collider energies.
\begin{center}
\begin{tabular}{|c|c|c|c|}
\hline
\rule[-1.2ex]{0mm}{4ex}$\sqrt{s}$ (GeV) & $2 f_2^R$  &
$(f_1^R)^2$ & $(f_1^I)^2$ \\ \hline
155 & 0.056 &  0.039  & 0.001 \\
160 & 0.109 &  0.092  & 0.016 \\
165 & $-$0.056 &  0.055  & 0.083 \\
170 & $-$0.072 &  0.031  & 0.080 \\
\hline
\end{tabular}
\end{center}
We see that the real parts are numerically dominant at threshold,
whereas for $E \gg \Gamma_W$ there is a strong cancellation
between $(f_1^I)^2$ and $2 f_2^R$.

In Ref.~\cite{BARDIN} it was suggested that a reasonable
approximation to the higher-order Coulomb corrections could be
obtained by using the  stable particle expansion (\ref{stable})
with the `exact' first-order off-shell correction as the
expansion parameter. This corresponds to
\begin{equation}
\label{approxform}
\vert f\vert^2 \; \approx \;
 1 \; + \; {X \over 2} \; + \;
{X^2 \over 12} \; + \; \ldots
\end{equation}
with (see Eq.~(\ref{fnlo}))
\begin{equation}
\label{Xis}
 X\; = \;  {\alpha\sqrt{s}\over 2  p}
\left[ \pi -2\; \mbox{arctan}\left({\vert\kappa\vert^2 -p^2
\over
 2 p\; {\rm Re}(\kappa ) }  \right)\right]  \; .
\end{equation}
Fig.~2 shows the exact and approximate second-order corrections
as a function of $\sqrt{s}$. Formally, the two results must
become equal far above threshold, where both must coincide
with the stable-$W$ result.  This behaviour is evident for
$\sqrt{s} \gapp 170$ GeV.  However, the approximation
clearly breaks down around and below threshold as expected (for
example, by comparing Eqs.~(\ref{threshold}) and
(\ref{approxform})).

\section{Conclusions}
\label{sec:conc}
In this note we have presented analytic and numerical
results for the first- and second-order Coulomb corrections
to the $e^+e^- \to W^+W^-$ cross section in the threshold region.
In fact the corrections are known to {\it all} orders,
see Eq.~(\ref{fdef}),
although it is clear from Fig.~1 that in practice the
first- and second-corrections are sufficient for  phenomenology
at the LEP2 collider. In terms of determining $M_W$ from a
precision
measurement of the threshold cross section, the inclusion of the
first-
and second-order Coulomb corrections is equivalent to a shift in
$M_W$ of 100~MeV and  3.4~MeV respectively.

We have also derived analytic expressions for the corrections
far above, far below, and close to threshold. For the former,
the well-known stable-$W$ results are reproduced. For the latter,
we have shown that the effective expansion parameter right at
threshold
is $\alpha \pi \sqrt{M_W/\Gamma_W} \approx \textstyle{1\over 7}$
rather than $\alpha \approx \textstyle{1 \over 137}$, and this
explains
the overall size of the first- and second-order corrections in
this
region.

Finally, we have studied the validity of the approximation
\cite{BARDIN} in which
the stable-$W$ all-orders result is combined with the
unstable-$W$
first-order result.  As expected the approximation works well
away from threshold, but is seen to break down for $\sqrt{s}
\lapp 165$ GeV.

\section*{\Large\bf Acknowledgements}

\noindent WJS and VAK are grateful to the UK PPARC for support.
This work was supported in part by the U.S.\ Department of
Energy,
under grant DE-FG02-91ER40685 and by the EU Programme
``Human Capital and Mobility'', Network ``Physics at High Energy
Colliders'', contract CHRX-CT93-0319 (DG 12 COMA).
\goodbreak

\section*{Figure Captions}
\begin{itemize}

\item [{[1]}]
The $O(\alpha)$ and $O(\alpha^2)$ Coulomb corrections of
Eq.~(\ref{sigcoul})
normalized to the Born cross section, as a function of the
$e^+e^-$ collider
energy $\sqrt{s}$.  The position of the nominal threshold
$\sqrt{s} =
2 M_W$ is marked.

\item [{[2]}]
The $O(\alpha^2)$ Coulomb correction of
Fig.~1 (solid line) compared to the approximate form of
Eqs.~(\ref{approxform},\ref{Xis})  \cite{BARDIN} (dashed line).

\end{itemize}
\newpage

\begin{figure}[h]
\begin{center}
\end{center}
\label{1}
\end{figure}
\newpage

\begin{figure}[h]
\begin{center}
\end{center}
\label{2}
\end{figure}

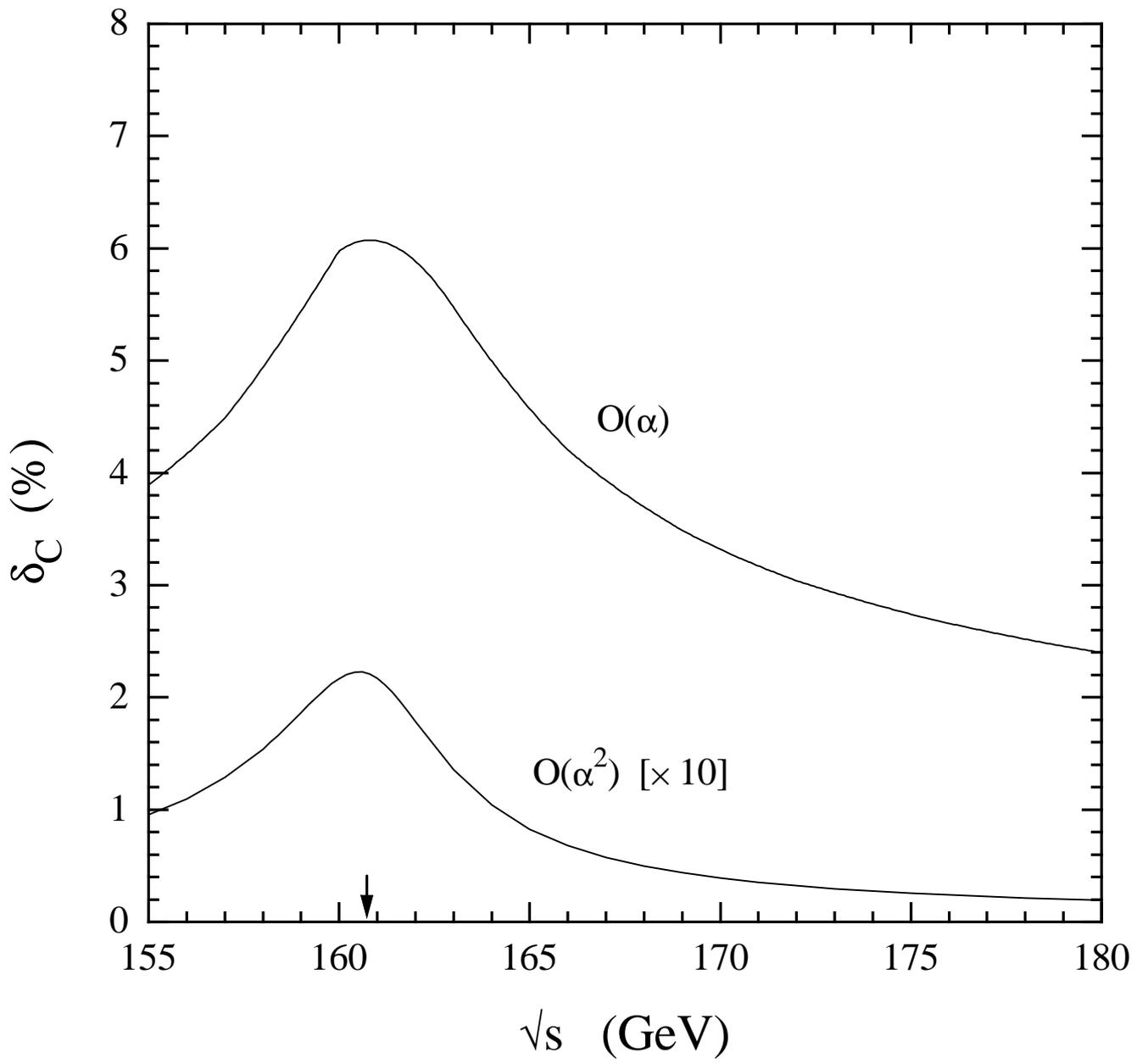
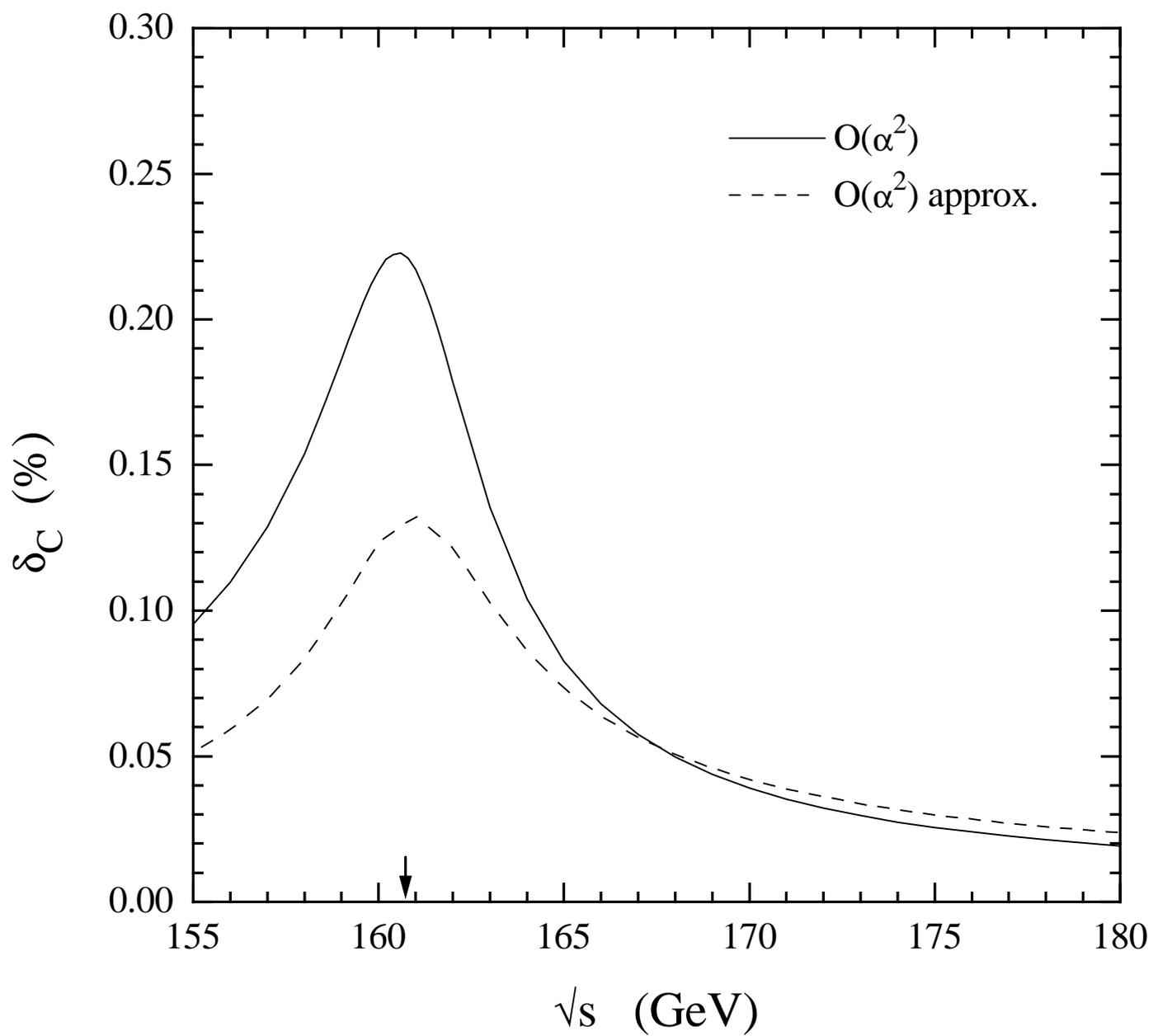
\end{document}